\documentclass{aa}
\usepackage{epsf,rotate}
\begin{document}
%
%   \thesaurus{ 06
%              (08.09.2 RW UMi;   % Stars: individual
%               08.14.2;          % novae, cataclysmic variables
%              08.02.1           % (stars:) binaries(including multiple): close
%               02.01.2           % Accretion, accretion discs --
%               08.18.1           % Stars: rotation
%               08.23.1)          % white dwarfs
%   }

   \title{The detection of a 1.4-h period in RW Ursa Minoris -- candidate 
for shortest recorded orbital period nova}

%    \subtitle{I. Overviewing the $\kappa$-mechanism}

\author{
A.~Retter$^{1}$ and Y. Lipkin$^{2}$\\
$^1$Dept. of Physics, Keele University, Keele, Staffordshire, 
ST5 5BG, U.K.; \\
$^2$School of Physics and Astronomy and the Wise Observatory,
Raymond and Beverly Sackler Faculty of Exact Sciences,
Tel-Aviv University, Tel Aviv, 69978, Israel; \\
}

%   \author{G. Wuchterl
%          \inst{1}
%          \and
%          C. Ptolemy\inst{2}\fnmsep\thanks{Just to show the usage
%          of the elements in the author field}
%          }

   \offprints{A.~Retter (ar@astro.keele.ac.uk)}

%   \institute{Institute for Astronomy (IfA), University of Vienna,
%              T\"urkenschanzstrasse 17, A-1180 Vienna\\
%              \email{wuchterl@amok.ast.univie.ac.at}
%         \and
%             University of Alexandria, Department of Geography, ...\\
%             \email{c.ptolemy@hipparch.uheaven.space}
%             \thanks{The university of heaven temporarily does not
%                     accept e-mails}
%             }

   \date{Received 9 August 2000; accepted 19 October 2000}
%   \date{Received September 15, 1996; accepted March 16, 1997}

\authorrunning {Retter \& Lipkin}

\titlerunning {1.4-h period in RW UMi}

% \begin{titlerunning} 
% Retter \& Lipkin -- 1.4-h period in RW UMi
% \end{titlerunning} 

% \begin{abstract}

\abstract{
CCD photometry of the classical nova RW~UMi during 10 nights in 1995
through an $R$ filter and 14 nights in 1997 through a `clear' filter, 
with the 1-m Wise telescope yields a periodic modulation in the light 
curve with the period 0.05912$\pm$0.00015 d and a semi amplitude of 
0.025 mag. We discuss several explanations for the periodicity, and 
suggest that it is likely to be the orbital period of the binary system
or a superhump period. RW~UMi is thus a candidate for the classical nova 
with the shortest known orbital period.
}

\maketitle

\keywords{individual: RW UMi --
novae, cataclysmic variables --
binaries: close -- 
Accretion, accretion discs --
white dwarfs}
%       \keywords{giant planet formation --
%                 $\kappa$-mechanism --
%                 stability of gas spheres
%                }
% accretion, accretion discs -- novae -- stars: individuals: RW UMi 
%    \end{abstract}

%
%________________________________________________________________

\section{Introduction}

RW~UMi is a high galactic latitude nova that erupted in 1956. Its outburst 
was reported, however, only about six years later (Kukarkin 1962). The 
nova reached a maximum of at least $m_{V}$$\approx$$6.0$ on 1956 September 
24, and decayed to $m_{V}$$\approx$11.5 a year later. Kukarkin also 
concluded that the progenitor of the nova was fainter than $m_{V}$=21.2 in 
POSS. Duerbeck (1987), however, barely detected the pre-nova at $m_{V}$=21.0 
in the same plate. Photometric observations of the object during the last 
two decades are consistent with a visual magnitude stable around 
$m_{V}$$\approx$18.8 (Cohen 1985; Kaluzny \& Chlebowski 1989; Szkody et al. 
1989; Howell et al. 1991; Szkody \& Howell 1992; Szkody 1994; Ringwald et 
al. 1996; Downes \& Duerbeck 2000). We note that the estimate of the nova 
in 1988 as $m_{V}$=21.0-21.5 (Howell 1988) was corrected by Szkody et al. 
(1989).

Spectra of the post-nova are presented by Kaluzny \& Chlebowski (1989), 
Szkody \& Howell (1992) and Ringwald et al. (1996). These show a strong 
blue continuum and several weak emission lines -- the Balmer and He II 
4686 lines. 

% Kaluzny \& Chlebowski found some splitting in the emission lines suggesting
% a relatively high inclination angle.

% -- an evidence for an accretion disc 

Cohen (1985) marginally detected a nebula around the remnant of RW~UMi 
using an image taken through a narrow H$\alpha$ filter. Slavin et al. 
(1995) repeated this observation, and concluded that the distance to the 
binary system is 5$\pm$2 Kpc. Esenoglu et al. (2000) resolved the shell 
of RW~UMi in 1995 and derived a distance range for the nova of 4200-5840 pc.
Downes \& Duerbeck (2000) failed, however, to detect the nova nebula in
an observation carried out in 1998.

Kaluzny \& Chlebowski (1989) found that the nova was not variable at the 
0.05 mag level in the $B$ and $V$ filters in five images taken in 1985 and 
one in 1987. Szkody et al. (1989) reported the discovery of a variation of 
about 0.2 mag in a 3-h observation in 1988 using a wide band red filter. 
They suggested a possible period of 117$\pm$5 minutes. Shafter \& Campbell 
(1990) independently announced a similar detection. Howell et al. (1991) 
observed the nova with a $B$ filter for 2.1 h in 1989, and proposeed a 
period of 113$\pm$10 minutes.

There are only three recorded classical novae with established orbital 
periods below the cataclysmic variables (CVs) period gap -- CP~Pup 1942, 
GQ~Mus 1983 and V1974~Cyg 1992 (Diaz \& Bruch 1997; Ritter \& Kolb 
1998\footnote{Note that the dwarf novae WX~Cet and TV~Crv were 
misclassified as novae.}). This is compared with about one hundred and twenty 
other short orbital period systems. In comparison, above the gap there are 
more than thirty novae and about one hundred and thirty other systems. Thus 
the ratio of novae to all CVs below the gap is nearly ten times smaller than 
above the gap. An increase of the number of known novae below the period gap 
is, therefore, necessary for the understanding of nova cycles of short 
orbital period CVs. The possibility that the orbital period in RW~UMi is 
below the gap lead us to add this nova to our list of targets in an ongoing 
programme, maintained at the Wise Observatory, to search for periods in 
post-novae. Retter (1999) announced the detection of a 1.4-h periodicity in 
the light curve of RW~UMi. In this work we elaborate that report, discuss 
possible physical mechanisms for the variation and further try to classify 
the system within the known subgroups of CVs.

% In 1995, the object was used mainly to fill gaps in the observational 
% plan; in summer 1997 RW~UMi became our primary target during dark nights.

% Primary analysis of the light curve of the nova failed to show a consistent 
% periodicity (Retter \& Leibowitz 1998; Retter 1999a), however, further 
% inspection of the data, lead to positive results announced by Retter 
% (1999b). 

\section{Observations}

We observed Nova RW~UMi during 10 nights in 1995 through a standard 
Cousins $R$ filter and 14 nights in 1997 through a `clear' filter. A few 
images of the nova in the $V$, $R$ and $I$ filters were taken as well in 
1995 May 11. In addition, one snapshot per night through the $R$ filter 
was taken on 98 occasions between 1997 October and 1999 July. Table~1 
presents a summary of the observation schedule. The photometry was carried 
out using the 1-m telescope at the Wise Observatory, with the Tektronix 
1K CCD camera. The typical exposures times were 3-15 m in 1995 and 5-6 m 
in 1997-1999. 

% The $R$ magnitudes of RW~UMi, measured in ?????? using the comparison 
% star ?????  is: $m_{R}$=18.5$\pm0.2$.

\begin{table} \label{table:v1500refs}
%\caption{The flux in the orbital modulation of V1500 Cyg.}
\caption[]{The observations time table}
\begin{center}
%\begin{tabular}{lcc}
\begin{tabular}{ccccccc}
%\begin{tabular}{@{}lcllccc@{}}

UT     & Time of  & Run  & Number & Filter & notes  \\
Date   & Start    & Time & of     &        &        \\
       & (HJD-    & (h)  & frames &        &        \\
       & 2449000) &      &        &        &        \\
\\

110595 & 849.275  &  0.9  & 2,6,1 & R,V,I    &          \\
%240595 & 862.312  &  1.9  & 10    & R        & clouds   \\
250595 & 863.259  &  2.9  & 17    & R        &          \\
060695 & 875.265  &  3.2  & 20    & R        &          \\
090695 & 878.254  &  1.3  & 8     & R        &          \\
100695 & 879.247  &  1.2  & 8     & R        &          \\
110695 & 880.238  &  1.6  & 10    & R        &          \\
180695 & 887.245  &  7.7  & 33    & R        & gaps     \\
200695 & 889.239  &  1.6  & 10    & R        &          \\
210695 & 890.234  &  1.4  & 9     & R        &          \\

020795 & 901.271  &  7.2  & 117   & R        &          \\
030795 & 902.252  &  7.5  & 105   & R        & gap      \\

130597 & 1582.235 &  7.9  & 78    & Clear    &          \\
140597 & 1583.308 &  5.2  & 48    & Clear    &          \\
100697 & 1610.251 &  7.6  & 86    & Clear    &          \\
110697 & 1611.251 &  7.6  & 80    & Clear    &          \\
120697 & 1612.253 &  7.6  & 85    & Clear    &          \\
270697 & 1627.253 &  7.5  & 65    & Clear    & gaps     \\
070897 & 1668.255 &  7.3  & 68    & Clear    &          \\
080897 & 1669.252 &  7.0  & 65    & Clear    &          \\
090897 & 1670.242 &  6.8  & 54    & Clear    &          \\
270897 & 1688.223 &  3.3  & 32    & Clear    &          \\
280897 & 1689.236 &  8.0  & 76    & Clear    &          \\
310897 & 1692.235 &  7.2  & 62    & Clear    &          \\
010997 & 1693.225 &  4.0  & 36    & Clear    &          \\
020997 & 1694.213 &  8.3  & 71    & Clear    &          \\

101097-& 1732-    &       &       &          &           \\
270799 & 2387     &       & 98    & R        & snapshots \\

%320  &$ 1060 \pm 200 $& Kemp, Sykes \&                \\
 &$ $& 

\end{tabular}
\end{center}
%$^\dag$These points are from spectroscopy, band pass 4750-4790\AA.
\end{table}

Bias and flat field corrections were made to each frame. Aperture 
photometric measurements were carried out using the DAOPHOT program 
(Stetson 1987). Differential magnitudes of the nova relative to several 
reference stars (the exact number depending on each image quality) were 
obtained from the images. The Wise Observatory reduction program DAOSTAT 
(Netzer et al. 1996) was used to reject variable comparison stars. The 
number of frames obtained in each filter on our programme is: 437 ($R$), 
906 (`clear'), 6 ($V$) and 1 ($I$). The mean of the errors in the 
observations in the $R$ and `clear' filters were 0.11 and 0.04 mag 
correspondingly.

% The measurements taken in 1995 May 24, were rejected due to the presence 
% of clouds during the observations.

% The mean of the errors in the continuous observations in 1997 through the 
% `clear' filter was 0.041 mag and with a standard deviation of 0.012. For 
% the snapshots in 1997-1999 the values were 0.10 and 0.04 correspondingly.

Fig.~1 presents all observations in the $R$ and `clear' bands during 
1995-1999. The two filters were combined assuming that the mean `clear'
magnitude of the comparison stars is the same as the mean $R$ magnitude.
The graph shows that the colour $R$-`clear' of the nova is relatively small 
compared with the scatter of the points that cannot be attributed only to 
the observational errors. The nightly means in 1997 varied by a few tenths 
mag from night to night with a standard deviation of $\sigma$=0.09 mag -- 
more than twice the mean error of the data points. During the first six 
nights in 1997 the nova appeared to be consistently about 0.1 mag brighter 
than the remaining eight nights.

%                                                One column figure
%----------------------------------------------------------- S_vib
   \begin{figure}
%      \vspace{5cm}
% \rotate[l]{\epsfxsize=60mm
{\epsfxsize=88mm
\epsffile{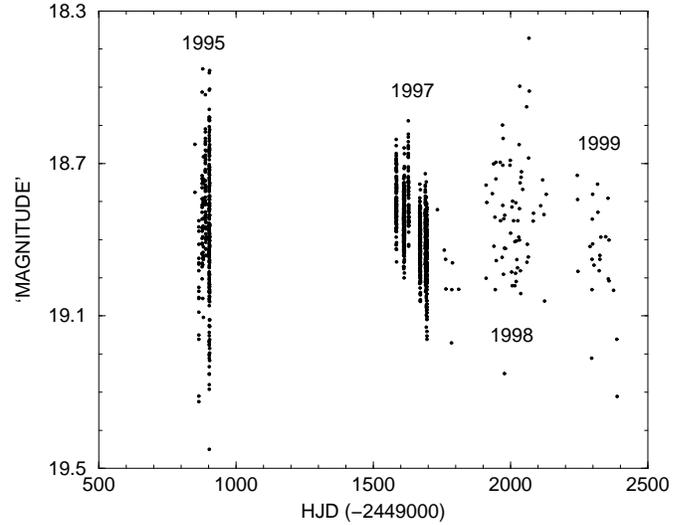}}
%\vspace{0cm}
%\hspace{0cm}\psfig{figure=fig1.eps,width=8.8cm}
%\vspace{0cm}

\caption[]{All 1995-1999 data points are presented. Note that this is a 
combination of two filters ($R$ and `clear'). In 1995 and 1997 RW~UMi was 
observed for a few hours each night; from winter 1997 on only snapshots 
were taken. In 1997 the object faded by about 0.1 mag during the second 
part of the observations (August-September) relative to the first part 
(May-June).}

\label{FigVibStab}

\end{figure}

\section{Data Analysis}

The best nights in the 1995 data showed a possible variation of the order 
of 1-3 h, however, since an $R$ filter was used, the errors were much 
larger than the 1997 data when more photons were accumulated using a 
`clear' filter. In addition, the length of run on each night in 1995 was 
generally very short and two out of the three longest nights have gaps 
in between. Therefore, the power spectrum of these data is very noisy.

The 1997 data were obtained without a filter, thus having smaller errors. 
The mean length of the runs was 6.8 h. A variation of $\sim$1.4 h with a 
full amplitude of 0.1-0.2 mag can be seen by simple visual inspection of 
the light curve in a few of the best nights. Fig.~2 presents two examples 
-- June 12 and August 7. In both light curves more than five cycles of the 
candidate periodicity are easily seen. Note the overall fading during the 
second night. Such gradients appeared in the light curves of several other 
nights as well.

% The 1997 light curve is shown in Fig.~2.

% \begin{figure}

% \centerline{\epsfxsize=3.0in\epsfbox{fig2.eps}}
% {\epsfxsize=88mm
% \epsffile{fig2.eps}}

% \caption[]{The 1997 data (`clear' filter). Empty circles show the 
% nightly means. Note that the object faded by about 0.1 mag during the 
% second part of the observations (August-September) relative to the 
% first part (May-June).}

% \end{figure}

\begin{figure}

% \centerline{\epsfxsize=3.0in\epsfbox{fig3.eps}}

{\epsfxsize=88mm
\epsffile{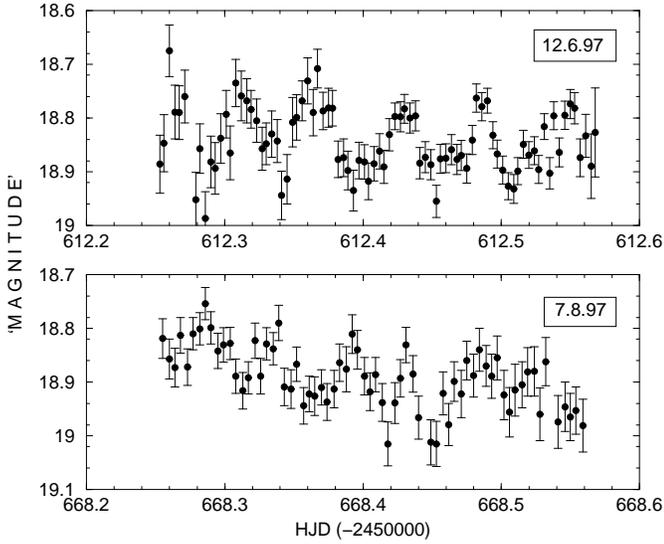}}

\caption[]{Light curves of two of our best nights in 1997 obtained using 
a `clear' filter. The upper panel shows the points of June 12 and lower 
panel -- August 7. More than five cycles of the 1.4-h periodicity can be 
seen in each night.}

\end{figure}

The power spectrum of the 1997 data obtained through the `clear' filter 
is shown in Fig.~3. The detrending was carried out by subtracting the 
mean from each night. The highest peak at mid-frequencies (f=16.916 d$^{-1}$)
corresponds to a period of 0.05912 d. It is many standard deviations 
above the noise level. Aliases of this peak are also present. To check 
whether the second highest peak at mid-frequencies after the 16.916 d$^{-1}$ 
frequency, its 1-d alias at 17.916 d$^{-1}$, is the correct period, we 
carried out the following test. First a synthetic light curve was built 
using a sinusoid at this period. This sinusoid was given the same amplitude 
it has in the data and sampled according to the window function. Next noise 
was added using the rms of the residuals of the fit to the data. We then 
checked whether the highest peak in a small interval around the 16.916 
d$^{-1}$ frequency is higher than the strongest peak near the 17.916 d$^{-1}$ 
frequency. One thousand simulations show that the probability that the 1-d 
alias at 17.916 d$^{-1}$ (0.05582 d) is the correct period is 4\%. 

The group of peaks at the left-hand side of the diagram is centered around 
the 2.0 d$^{-1}$ peak that appears in the window function as well. Other 
low-frequency peaks correspond to periodicities that are longer than the 
typical interval of observations in each night, and probably reflect the 
strong noise level and / or gradients that appear in a few individual nights. 
A consistent search for a second periodicity in the light curve with the 
various techniques discussed by Retter et al. (1997) yielded no positive 
results. We note that the power spectrum of the data after the subtraction 
of the linear trend from each night removes most low-frequency peaks, but 
the structure around the 17.916 d$^{-1}$ frequency is quite similar to Fig.~3.

% 0.05912$\pm$0.00015 d.
% The other peaks seen in the graph around the highest peak are identified 
% as aliases.

\begin{figure}

%\centerline{\epsfxsize=3.0in\epsfbox{fig4.eps}}

{\epsfxsize=88mm
\epsffile{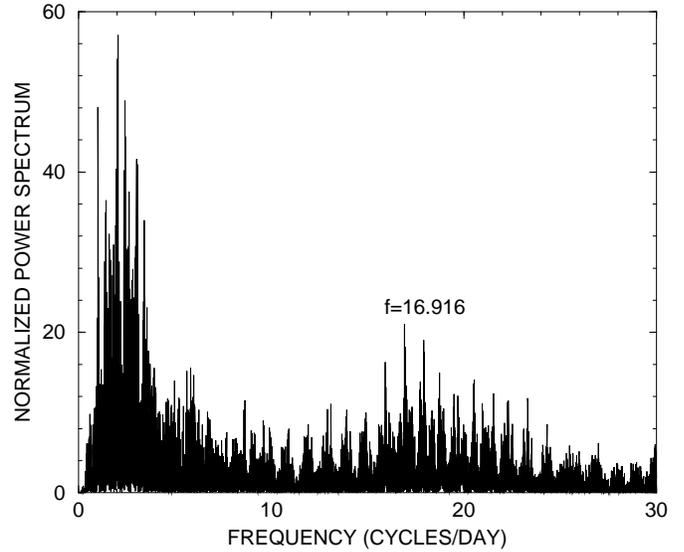}}

\caption[]{The power spectrum of the 1997 points (`clear filter') after
the subtraction of the nightly means. The peak at the frequency 16.916 
d$^{-1}$, marked as f, corresponds to the period 0.05912 d. The group of 
peaks at the left-hand side of the diagram correspond to 1-d, 0.5-d...
aliases that also appear in the window function and to frequencies longer 
than the length of the observations in each night, which are probably the 
result of the noise, the detrending method used and the presence of nightly 
gradients.}

\end{figure}

In Fig.~4 we present the mean light curve of the 1997 observations. The 
nightly trends were removed by subtracting a linear term from each night. 
The light curve shows a sinusoidal variation. The first harmonic fit to 
the data yielded a semi amplitude of 0.025$\pm$0.006. The error corresponds 
to a 99\% confidence level, and it was calculated by a sample of 1000 
bootstrap simulations (Efron \& Tibshirani 1993). The best fit ephemeris 
for the periodicity is:\\\\
T${min}$ = HJD 2450612.397+0.05912 E \\
\hspace*{1.2in}$\pm$0.001$\pm$0.00015 (2.5$\sigma$)\\

%\hspace*{1.1in}$\pm $0.0005 \hspace*{0.01in}  $\pm$ 0.000001\\

\begin{figure}

% \centerline{\epsfxsize=3.0in\epsfbox{fig5.eps}}

{\epsfxsize=88mm
\epsffile{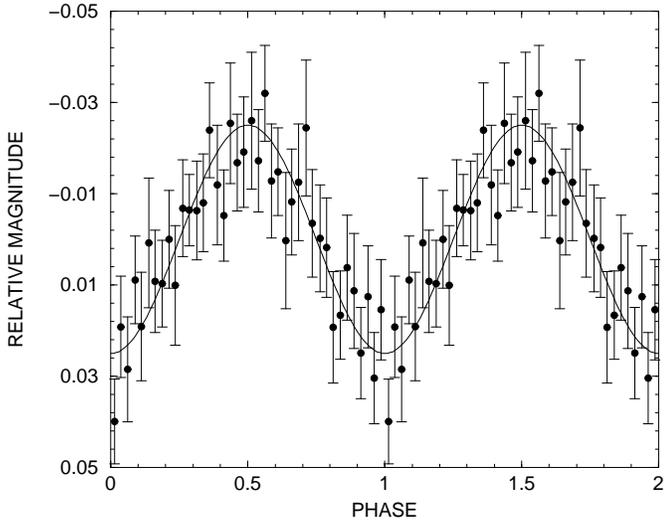}}

\caption[]{The `clear' filter light curve of the nights in 1997 folded on
the 1.4-h period and binned into 40 equal bins. The solid line represent
the sinusoidal fit to the data. The bars show the standard error derived
from the rms within each bin.}

\end{figure}

\section{Discussion}

% \subsection{Interpretation of the period and classification of the system --
% general remarks}

The light curve of RW~UMi in 1997 was modulated by the 1.4-h period. There 
are three obvious interpretations for this periodicity. It might represent 
the spin period of the white dwarf, a superhump variation or the orbital 
period of the binary system. In the following we examine the different 
models and try to classify the system in the light of the observations 
accumulated so far on the nova.

\subsection{A spin period-?}

First, we discuss the question whether the system is an AM~Her (polar) 
system (for a review see Warner 1995). The primary white dwarfs in these 
systems possess strong magnetic fields. Thus, the accreted matter from the 
companion star flows directly to the two poles of the rotating white dwarf 
whose spin period is synchronized with the orbital period of the binary 
system.

% the question whether the system is magnetic or non-magnetic

The spectra of RW~UMi show a strong continuum and only weak emission 
lines. There are no strong high excitation lines, and the He II 4686 line 
is rather weak (Sect.~1). We thus believe that the polar scenario can be 
eliminated, since typical spectra of AM~Her systems have a weak continuum
and strong emission lines (Warner 1995). The typical optical spectra, 
radiated from accretion discs in post-novae, are characterized by a strong 
continuum, because the disc is usually optically thick in classical nova 
systems for a few decades after their eruptions (Retter et al. 1999).

If the magnetic field is moderate the spin period is usually much shorter 
than the orbital period and an accretion disc might be formed around the 
white dwarf. These objects are termed `intermediate polars' (for reviews see 
Patterson 1994; Hellier 1996). Spin periods in intermediate polars range 
between 33 s in AE Aqr (Hellier 1996) to 1.44 h in Nova V1425~Aql 1995 
(Retter et al. 1998). RW~UMi might thus be an intermediate polar with the 
1.4-h period being the rotation period of the white dwarf. The orbital 
period should then be somewhat longer than the spin period. The power 
spectrum of RW~UMi (Fig.~3) does not show, however, other periodicities in 
addition to the 1.4-h period (Sect.~3). Therefore, it seems unlikely that 
the periodicity we have discovered is a spin period. 

% Our observations thus seem to reject the presence of any long term 
% periodicity. 

\begin{table*} \label{table:v1500refs}
%\caption{The flux in the orbital modulation of V1500 Cyg.}
%\caption[]{The observations time table}
\caption[]{Properties of the classical novae below the period gap}
\begin{center}
%\begin{tabular}{lcc}
\begin{tabular}{lcllccc}

Object name&year of outburst&orbital period& nature         &pre-nova magnitude&post-nova magnitude&difference\\

\\

RW~UMi     & 1956   & 1.419-?$^{1}$  & non-magnetic-?$^{1}$ & $21^{2}$         & 18.8$^{1}$        & 2.2      \\
GQ~Mus     & 1983   & 1.425$^{3,4}$  & magnetic-?$^{3,4}$   & $22^{2}$         & 18.0-?$^{4,5}$    & 4-?      \\
CP~Pup     & 1942   & 1.474$^{6}$    & non-magnetic$^{6}$   & $>$$17^{7}$        & 15.3$^{6}$        & $>$$1.7$   \\
V1974~Cyg  & 1992   & 1.950$^{8,9}$  & non-magnetic$^{8,9}$ & 21$^{8,10}$      & 16.5--?$^{11}$    & 4.5--?   \\

\end{tabular}
\end{center}

%$^\dag$These points are from spectroscopy, band pass 4750-4790\AA.

$^1$This work;
$^2$Duerbeck 1987; 
$^3$Diaz \& Steiner 1989;
$^4$Diaz \& Steiner 1994;
$^5$Diaz et al. 1995;
$^6$Patterson \& Warner 1998;
$^7$Warner 1995;
$^8$Retter \& Leibowitz 1998;
$^9$Skillman et al. 1997;
$^{10}$Pavelin et al. 1993;
$^{11}$Goransky 2000.

\end{table*}

%___________________________________ Two column table (place early!)
%   \begin{table*}
%      \caption[]{Properties of Classical novae below the period gap}
%         \label{TabSecInst}
%      \vspace{4cm}
% \item[$^{\mathrm{a}}$] This is footnote a
%   \end{table*}

\subsection{A superhump period-?}

Superhumps are quasi-periodic oscillations that have been observed in the
light curves of non-magnetic CVs with relatively short orbital periods 
(typically $P_{orbital}$$<$4 h). They either appear in the bright outbursts 
(superoutbursts) of the SU~UMa subclass of dwarf novae or permanently in 
nova-likes. Superhump periods differ from binary periods by a few percent. 
They are understood as the beat periods of the orbital period and the 
apsidal (or the nodal) precession period of the accretion disc. For 
reviews on superhumps see Warner (1995) and Osaki (1996) and on permanent 
superhumps -- Patterson (1999) and Retter \& Naylor (2000). If the 1.42-h 
period in RW~UMi is a positive superhump, a superhump period excess of 
1-4\% is expected according to the observed values in superhump systems 
(Ritter \& Kolb 1998), leading to a binary period in the range 1.36-1.40 h. 
If the periodicity is a negative superhump, the expected range in the binary 
period is 1.43-1.45 h since period deficits in negative superhumps are 
about half period excesses in positive superhumps (Patterson 1999). In both 
cases the orbital period is only a few percent different from the observed 
periodicity.

\subsection{The binary period-?}

Orbital periods of nova systems range between 1.425 h (GQ~Mus) to 48 h 
(GK~Per) with a peak around 3-4 h (Diaz \& Bruch 1997). The observed 
orbital period distribution of CVs presents a deficiency of systems 
between about 2 h and 3 h, which is termed 'the period gap' (Warner 1995). 
So far only three novae with established orbital periods below the gap have 
been discovered. This is compared with a few dozen novae with orbital 
periods above the gap (Sect.~1). If the 1.419-h periodicity observed 
in the light curve of RW~UMi is its orbital period, it would be the 
shortest recorded binary period in a classical nova, although confirmation by 
radial velocity measurements is required. Table~1 presents the pre-nova 
and post-outburst magnitudes of the novae below the gap as well as several 
other properties. The data show that the brightness of the post-novae 
below the gap is still above the pre-nova state. RW~UMi is much brighter 
than the progenitor value as well.

% $P_{orb}$=1.425 (

% The detected periodicity in the optical light curve of RW~UMi is 
% slightly shorter than the orbital period in GQ~Mus (see Table~2). 

\subsection{RW~UMi as a high galactic latitude nova}

RW~UMi is a high galactic latitude nova and therefore probably a binary 
system in the halo of the galaxy. Halo CVs are believed to be relatively 
old systems compared with the population of CVs in the galactic disc (see
e.g. Howell \& Szkody 1990). Thus, according to evolutionary theories halo 
systems should have shorter orbital periods (assuming they haven't bounced 
yet at the period minimum). Such a difference between the two populations 
has indeed been found (Howell \& Szkody 1990). The possibility that the 
orbital period of RW~UMi is the shortest among all known novae is 
consistent with this idea.

\subsection{The mass accretion rate and thermal stability of RW~UMi}

In this section we estimate the mass transfer rate in RW~UMi to assess 
whether it has dwarf nova or nova-like characteristics. Permanent 
superhumpers (and nova-likes) are believed to occur in CVs with thermally 
stable discs, while dwarf novae (including SU~UMa systems) are thermally 
unstable. Based on the thermal-tidal disc instability model (Osaki 1996), 
assuming that the disc is the dominant light source in the visual band, 
and neglecting a possible contribution from the hot white dwarf in 
post-novae, Retter \& Leibowitz (1998) and Retter \& Naylor (2000) 
developed a way to check the thermal stability state of disc-accreting 
CVs. In the following, we apply this method to RW~UMi to obtain its mass 
transfer rate. 

The visual magnitude of the post-nova is $m_{V}$$\approx$18.8 and the 
distance -- 4200-5840 pc (Sect.~1). The interstellar reddening is 
$A_{V}$=0.1 (Anhert 1963; Cohen 1985). For the mass of the white dwarf, 
we take $M_{WD}$=0.7$M_{\odot}$ -- typical to CVs below the gap (Patterson 
1998). The absence of eclipses (Sect.~3) limits the inclination angle to 
i=0-65$^{\circ}$. Inserting all these values into eq.(3) of Retter 
\& Naylor (2000), we obtain $\dot{M}$=1.2-8.6$\times 10^{17}$g s$^{-1}$. 
The critical mass transfer rate, calculated by eq.(2) of Retter \& 
Leibowitz (1998), assuming that the orbital period is 1.4 h (Sect.~4.3) 
is 0.5$\times 10^{17}$g s$^{-1}$. The system is thus very likely to be 
thermally stable. This fact together with the long-term light curve of the 
nova (Fig.~1) that does not show dwarf nova outbursts (although obviously
outbursts could have been missed) seem to reject the possibility that the 
post-nova is a dwarf nova system, and therefore it should be a nova-like 
system. A similar test applied to the progenitor of the nova at $m_{V}$=21.0 
(Sect.~1) yields $\dot{M}$=0.2-1.1$\times 10^{17}$g s$^{-1}$. This range 
straddles the limiting value, so the thermal stability state of the 
pre-nova cannot be determined.

\subsection{Final classification of the system}

A definite classification of the system cannot be made. An AM~Her model 
doesn't fit the nova properties (Sect.~4.1). An intermediate polar model 
might be applicable to RW~UMi, but only if an additional short-term 
periodicity (the spin period) or a relatively long period (which would be 
the orbital period, making the 1.4-h period the spin period) is found in 
the future. We note that it is very likely that an accretion disc is 
formed around the white dwarf in RW~UMi as the AM~Her scenario has been 
rejected and since it is believed that most intermediate polars have 
accretion discs (Patterson 1994).

A dwarf nova (SU~UMa) model seems to be ruled out by the long-term behaviour
of the light curve in 1995-1999 (Fig.~1) that does not show any dwarf nova 
outburst, and by our findings that the system is almost certainly thermally 
stable (previous section). RW~UMi is therefore most likely a system with a
short orbital period in a high state. As such, it should have permanent 
superhumps according to Osaki (1996), however, the observations accumulated 
so far cannot confirm this phenomenon, probably because the nova is 
relatively faint.

% \subsection{RW~UMi as the shortest orbital period nova}

\section{Summary}

A photometric period of 1.4-h was detected in the light curve of the 
classical nova RW~UMi. We argue that it is most likely the orbital period 
of the system or a superhump period. In both models Nova Ursa Minoris 1956 
has a very short binary period. If confirmed, RW~UMi will be then the 
fourth nova below the gap and the third to be non-magnetic.

\begin{acknowledgements}

We thank the referee, Daisaku Nogami, for many valuable remarks.
We are grateful to Hasan Esenoglu who re-raised our interest in this 
object. Tim Naylor is acknowledged for a careful reading of the 
manuscript and for many useful comments. We also thank Coel Hellier and 
Sandi M. Catalan for several fruitful discussions, and John Dan and the 
Wise Observatory staff for their expert assistance with the observations.
AR is supported by PPARC. Astronomy at the Wise Observatory is supported 
by grants from the Israeli Academy of Sciences.

\end{acknowledgements}

\end{document}